\documentclass[12pt,a4paper]{article}

\usepackage{latexsym, amsfonts,amsmath,amssymb,graphics,bbm}
\usepackage{pifont}    
\usepackage{dsfont}
\usepackage{feynmp}    
\usepackage{nccmath}   
\usepackage[dvips]{graphicx}

\usepackage{mathrsfs}  
\usepackage{bbm} 
\usepackage{slashed}   
\DeclareMathAlphabet{\mathpzc}{OT1}{pzc}{m}{it}

\newcommand{\vphi}{{\pmb{\phi}}}

\newcommand{\rmt}{{\scriptscriptstyle\top}} 

\newcommand{\kA}[0]{}
\newcommand{\kB}[0]{}

\newcommand{\koment}[1]{}      
\newcommand{\nothing}[1]{}

\newcommand{\anti}[1]{\overline{#1}}

\newcommand{\U}[2]{\cU^{#1}_{\ \, #2}}

\newcommand{\derf}[2]{\frac{\delta #1}{\delta #2}}     

\newcommand{\dd}[0]{{\rm d}}

\newcommand{\pa}[0]{{\partial}}

\newcommand{\hb}{\hbar}
\newcommand{\id}[0]{\mathds{1}}

\newcommand{\refer}[1]{(\ref{#1})}

\newcommand{\nn}[0]{\nonumber}

\newcommand{\tG}[0]{{\widetilde{G}}}

\newcommand{\cA}[0]{{\mathcal{A}}}

\newcommand{\cF}[0]{{\mathcal{F}}}
\newcommand{\cG}[0]{{\mathcal{G}}}

\newcommand{\cI}[0]{{\mathcal{I}}}

\newcommand{\cL}[0]{{\mathcal{L}}}

\newcommand{\cO}[0]{{\mathcal{O}}}

\newcommand{\cT}[0]{{\mathcal{T}}}
\newcommand{\cU}[0]{{\mathcal{U}}}

\newcommand{\cW}[0]{{\mathcal{W}}}

\newcommand{\sL}[0]{{\mathscr{L}}}

\newcommand{\sX}[0]{{\mathscr{X}}}

\newcommand{\vA}[0]{{\mathbf{A}}}

\newcommand{\bA}[0]{{\mathbb{A}}}

\newcommand{\al}{\alpha}
\newcommand{\be}{\beta}
\newcommand{\de}{\delta}
\newcommand{\De}{\Delta}
\newcommand{\ga}{\gamma}
\newcommand{\Ga}{\Gamma}
\newcommand{\si}{\sigma}

\newcommand{\ka}{\kappa}
\newcommand{\la}{\lambda}
\newcommand{\La}{\Lambda}
\newcommand{\Th}{\Theta}
\newcommand{\ep}{\epsilon}

\newcommand{\cphi}{\varphi}

\newcommand{\om}{\omega}
\newcommand{\Om}{\Omega}
\newcommand{\ze}{\zeta}

\newcommand{\TS}[0]{\mathcal{T}}       

\newcommand{\volel}[1]{\!\! {{\rm d}^4 #1 }}

\newcommand{\matri}[1]{\left[#1\right]}

\begin{document}



\pagestyle{empty}
\begin{flushright}
\end{flushright}
\vspace*{5mm}

\koment{COMMENT!!!}

\begin{center}

{\Large {\bf 
Vector-scalar mixing to all orders}\\
\large{\bf for an arbitrary gauge model\\
\ in the generic linear gauge}}
\vspace*{1cm}

{Adrian Lewandowski 
\footnote{Email: lewandowski.a.j@gmail.com}
}\\

\vspace{1cm}
{{\it
Albert Einstein Center for Fundamental Physics,\\ 
Institute for Theoretical Physics, University of Bern, \\
Sidlerstrasse 5, CH-3012 Bern, Switzerland 
}}

\kB

\vspace*{1.7cm}
{\bf Abstract}
\end{center}
\vspace*{5mm}
\noindent
I give explicit fromulae for full propagators of vector and scalar fields 
in a generic spin-1 gauge model quantized in an arbitrary linear covariant 
gauge. The propagators, expressed in terms of all-order one-particle-irreducible 
correlation functions, have a remarkably simple form because of constraints originating from Slavnov-Taylor identities of Becchi-Rouet-Stora symmetry. 
I also determine the behavior of the propagators in the neighborhood of the poles,  
and give a simple prescription for the coefficients that generalize 
(to the case with an arbitrary vector-scalar mixing)  
the standard $\sqrt{\mathcal{Z}}$ factors of Lehmann, Symanzik and Zimmermann. 
So obtained generalized $\sqrt{\mathcal{Z}}$ factors, are indispensable 
to the correct extraction of physical amplitudes from the amputated correlation functions in the presence of mixing.

The standard $R_\xi$ guauges form a particularly important subclass of gauges considered in this paper. While the tree-level vector-scalar mixing is, by construction, absent in $R_\xi$ gauges, it unavoidably reappears at higher orders. Therefore the prescription for the generalized 
$\sqrt{\mathcal{Z}}$ factors given in this paper is directly relevant for the 
extraction of amplitudes in $R_\xi$ gauges.

\vspace*{1.0cm}


\vspace*{0.2cm}

\vfill\eject
\newpage

\pagestyle{plain}

\section{Introduction} 

Bosonic fields can mix with each other, unless 
the symmetries tell us otherwise. In the Standard Model (SM) of particle 
physics (see e.g. \cite{WeinT2}), there are four neutral elementary bosonic 
fields: 
scalar $h_0$, vector $\gamma_\mu$ (photon), vector $Z_\mu$ and (in renormalizable gauges) the scalar would-be Goldstone field $G_Z$ 
associated with $Z_\mu$. While the mixing between 
$h_0$ and the other fields requires a transfer of CP-violation from the quark 
sector (and therefore cannot appear at low orders of perturbation theory),  
there are dozens of SM extension in which $Z_\mu$ is mixed with  
physical scalars already at one-loop; 
the singlet Majoron model \cite{Majo} is perhaps the simplest extension of 
this sort. \footnote{ In fact, already in the SM there are nontrivial 
one-loop corrections to the photon-$Z_\mu$ mixing. The Landau-gauge description 
of this mixing (in the formalism used in this paper) can be found in 
\cite{ALLandau}. } 
Therefore, it would be welcomed 
to have a simple prescription 
that gives, in the presence of a generic mixing, 
the physical amplitudes directly it terms  
of amputated correlation functions. 
After all, in the case without mixing, there is a standard textbook algorithm, based on the 
$\sqrt{\mathcal{Z}}$ factors of Lehmann-Symanzik-Zimmermann (LSZ), 
which does the job in a simple and elegant manner, 
see e.g. \cite{Chank} for a nice 
description (with a derivation) in English. 

The standard first step in the LSZ-reduction \cite{Chank} 
is to study the behavior of propagators with resummed quantum corrections, 
and $\sqrt{\mathcal{Z}}$'s are square roots of the residues (up to the $i$ factor)  
of resummed propagators at the poles. However, despite many beautiful papers devoted 
to the study of physical (and unphysical) states in covariant gauges, of which I~specifically mention here Refs. \cite{BBBC,KugoOjima}, 
the complete and concise expressions for the resummed propagators of bosonic 
fields, in a generic gauge model and a generic linear covariant gauge, were 
not given in the literature, to the best of my knowledge. 
The present papers fills that gap, and completes the LSZ algorithm. 

In fact, the present paper is a completion of my two earlier papers 
\cite{ALFermions,ALLandau}, where the reader may find a more comprehensive 
list of relevant references. 
Inspired by the analysis of 2-by-2 and 3-by-3 mixing of 
Majorana fermions in  \cite{PilafRL1}, 
and almost generic (though complicated) analyses of n-by-n fermionic mixing in 
\cite{PlumOld} and \cite{Kniehl}, I gave in \cite{ALFermions}  a simple prescription for 
handling mixed fermions in a completely generic case, with no dependence 
whatsoever on particular renormalization conditions, including situations 
in which multiple states are associated with a single pole.
A generic prescription for dealing with mixed scalars in non-gauge 
theories was also given in \cite{ALFermions}.  
Next, in \cite{ALLandau} I provided a prescription for handling 
mixed vector-scalar systems in the Landau gauge. 

Before generalizing the results of \cite{ALLandau} to an arbitrary 
linear covariant gauge, in order to make the present paper as self-contained 
as possible,  I will recapitulate here the generalized LSZ algorithm 
in  purely scalar theories. 
Suppose that $\{\phi^j\}$ is a set of renormalized (in some convenient renormalization scheme) scalar fields.  Without loss of 
generality I assume that $\phi^j$ are Hermitian and have vanishing vacuum expectation values (VEVs). The renormalized one-particle-irreducible (1PI) correlation functions of scalars (with resummed quantum corrections) can be parametrized in 
the following way 
\eqs{\label{Eq:Gamma2-scalar} 
\widetilde{\Ga}_{k j}(-p,p)
&=&
S(p^2)_{kj} 
=
\Big[
p^2\id-M^2_S(p^2)
\Big]_{k j}
\,,
}
where $M^2_S(s)=M^2_S(s)^\rmt$ is a symmetric matrix. 
I emphasize here that all the relevant objects have been already renormalized, 
only because  
the procedure described below has, in principle, nothing to do with 
renormalization. In fact, the minimal subtraction schemes are by far the most 
popular ones these days, and (in this class of schemes) one always faces 
the problem of extracting physical results from renormalized correlation 
functions.

Inverting the matrix in 
Eq. \refer{Eq:Gamma2-scalar} we get the (connected) propagator
\eq{\label{Eq:G2-scalar} 
\widetilde{G}^{\,k j}(p,-p)
=
i 
\Big[ \big(p^2\id-M^2_S(p^2)\big)^{-1} \Big]^{k j}
\,,\
}
that has, as proved in \cite{ALFermions, ALLandau}, the following behavior about the poles  
\eq{\label{Eq:D-as-GENERAL-scalars-indi} 
\widetilde{G}^{kj}(p,-p)
=
\sum_\ell \sum_r
\, 
\tilde{\ze}^k_{S[\ell_r]} 
\,
\frac{i}{p^2-m^2_{S(\ell)}}
\,
\tilde{\ze}^j_{S[\ell_r]}
+\text{[non-pole part]}
\,.
}
It is clear that the pole masses $m_{S({\ell})}^2$ are solutions to the following 
equation
\eq{\label{Eq:GapEq-0-scalar}
\det(s\mathds{1}-M^2_S(s))\big|_{s=m_{S(\ell)}^2}=0\,.
}
It is also easy to believe that 
the coefficients $\tilde{\ze}_{S[\ell_1]}\,,\tilde{\ze}_{S[\ell_2]}\,,\ldots,$ form the basis of the corresponding eigenspace 
(in this notation $m_{S(\ell)}^2\neq m_{S(\ell')}^2$ for 
$\ell\neq\ell'$) 
\eq{\label{Eq:xi-Eig-GENERAL-scalars}
M^2_S(m_{S(\ell)}^2)\,  \tilde{\ze}_{S[\ell_r]} 
= 
m_{S(\ell)}^2\,
\tilde{\ze}_{S[\ell_r]}\,.
}
It turns out \cite{ALFermions, ALLandau} that  to prove the behavior 
\refer{Eq:D-as-GENERAL-scalars-indi}, one just need 
to find eigenvectors satisfying 
the following normalization/orthogonality conditions 
\eq{\label{Eq:norm-cond-GENERAL-scalars}
\tilde{\ze}_{S[\ell_r]}^{\,\, \rmt} \,  
\left[
\mathds{1} - M^2{}^{\prime}_{\!\!\!\!S\, } (m_{S(\ell)}^2)
\right]\,
\tilde{\ze}_{S[\ell _q]}
=\delta_{r q}\,, 
}
where 
$
M^2_S{}^{\prime}(s) 
\equiv 
\dd M^2_S(s)/\dd s\,.
$
\footnote{Note that the eigenvectors corresponding to different pole masses are, in general,  
not orthogonal to each other beyond the tree-level.}
Such eigenvectors always exist (in perturbation theory) for real and complex poles \cite{ALFermions, ALLandau}. 
Moreover, for real poles, one can always find eigenvectors 
$\tilde{\ze}^j_{S[\ell _q]}$ that obey \refer{Eq:norm-cond-GENERAL-scalars} 
and are themselves real.\footnote{ 
Of course, Eq. \refer{Eq:norm-cond-GENERAL-scalars} implicitly assumes 
that infrared divergences are absent, so that the 
derivative 
$M^2{}^{\prime}_{\!\!\!\!S\, } (m_{S(\ell)}^2)$ 
is finite. It is however worth saying that, even if certain 
matrix elements of $M^2{}^{\prime}_{\!\!\!\!S\, } (m_{S(\ell)}^2)$ 
are IR-divergent, 
one can still use the normalization conditions 
\refer{Eq:norm-cond-GENERAL-scalars} 
by replacing 
$M^2{}^{\prime}_{\!\!\!\!S\, } (m_{S(\ell)}^2)\mapsto M^2{}^{\prime}_{\!\!\!\!S\, } (q^2) $
and taking the $q^2\to m_{S(\ell)}^2$ limit, 
as long as limit of the left-hand-side of 
\refer{Eq:norm-cond-GENERAL-scalars} is finite 
(see Ref. \cite{ALLandau}). 
Such IR-divergences are therefore spurious 
and do not change the structure of the asymptotic states, 
in contrast to the ``physical" IR-divergences which are usually 
handled by introducing an IR-cutoff.
}

From the representation \refer{Eq:D-as-GENERAL-scalars-indi} 
of the propagator it is clear how to generalize the LSZ algorithm to the 
case of (purely scalar) mixing. Indeed, Eq. \refer{Eq:D-as-GENERAL-scalars-indi}
immediately shows that the asymptotic field $\vphi^j$ associated with 
the renormalized field $\phi^j$ has the following form
\eq{\label{Eq:AsymField}
\vphi^j= 
\sum_{  \ell } {}^\prime \sum_r
\tilde{{\zeta}}^j_{S[\ell_r]}  
\Phi^{\ell_r}\,,
}
where $\Phi^{\ell_r}$ are free Hermitian 
scalar fields  of mass $m_{S(\ell)}$ 
with canonically normalized 
propagators, and such that states created/annihilated by $\Phi^{\ell_r}$ and 
$\Phi^{\ell'_{r'}}\neq \Phi^{\ell_r} $ are orthogonal 
to each other. (Strictly speaking, asymptotic fields
$\Phi^{\ell_r}$ exist only for \emph{real} pole masses $m_{S(\ell)}$,
and therefore I put the prime on the first sum in \refer{Eq:AsymField}.) 
In other words, 
to obtain the correctly normalized (i.e. consistent with unitarity) 
amplitude of the process 
involving a particle corresponding to $\Phi^{\ell_r}$, one has to contract the 
eigenvector  $\tilde{\zeta}^j_{S[\ell_r]}$ with the amputated correlation functions 
$\cA_{j\ldots}(p,\ldots)$  of the renormalized scalar fields $\phi^j$ for 
$p^2=m^2_{S(\ell)}$.

Now I will explain how to generalize this algorithm to the mixing in gauge 
theories, beyond the Landau gauge, what is the main subject of this paper.

\section{Vector-scalar mixing in the generic case}

As before, I can assume 
that vector ($A^\al_\mu$) and scalar ($\phi^j$) fields are Hermitian. 
The gauge fixing Lagrangian in a generic linear covariant gauge 
has the following form \koment{p.Rxi1} \cite{BBBC} (see also \cite{PiguetSorella,WeinT2})
\eq{\label{Eq:LagrGaugeFix}
\cL_{\rm gauge-fix} 
 = 
h_\al f^\al + \frac{1}{2} \xi^{\al\be} h_\al h_\be\,, 
\qquad 
f^\al = - \pa^\mu\! A^\al_\mu - \U{\al}{j} (\phi^j + v^j), 
}
where $h_\al$ are Nakanishi-Lautrup fields (see e.g. \cite{WeinT2}),
\footnote{I apologize for a potentially confusing notation:  
$h_\al$'s have nothing in common with 
the Higgs field $h_0$ mentioned earlier.} 
while $\xi^{\al\be}$ and  $\U{\al}{j}$ are matrices of gauge fixing parameters 
($\xi^{\al\be}$ is symmetric). 
As before I assume that $\phi^j$ has a vanishing VEV, and therefore 
$v^j$ is an all-order VEV of the field $\phi^j+v^j$ 
in the ``symmetric phase". 
Note that (as is clear from, for instance, the path-integral representation), 
the quantum corrected propagators in the scalar-vector sector are independent 
of whether one decides to keep Nakanishi-Lautrup fields, or integrates them out. 
However, the corresponding 1PI functions do depend on this decision. 
Nonetheless, it is well known that there are no quantum corrections 
to the gauge fixing terms in linear gauges (see e.g. \cite{PiguetSorella}); 
in other words, the renormalized 1PI generating functional $\Gamma[\ ]$ 
to all orders of perturbation theory depends on $h_\al$ only through 
(the integral of) the tree-level gauge-fixing Lagrangian 
\refer{Eq:LagrGaugeFix}. 
Thus, the ``effective" 1PI two-point functions in the vector-scalar sector, 
obtained by integrating out $h_\al$'s, differ from the ``primordial" ones 
only by the terms originating from the ``effective" gauge-fixing Lagrangian 
\eq{\nn
\cL_{\rm gauge-fix} ^{eff}
 = 
- \frac{1}{2} {(\xi^{-1})}_{\al\be} f^\al f^\be \,. 
}
It turns out that the ``primordial" 1PI two-point functions are much closer 
to the physical reality (this fact is not particularly surprising, as 
they are independent of gauge-fixing parameters at the tree level), 
and therefore, in this paper, the $h_\al$'s are \emph{not} integrated out. 
The additional advantage of such an approach is that one has, at every intermediate 
stage, a nonsingular transition to the Landau gauge ($\xi=0=\cU$).

I use the following convention
\begin{eqnarray}
\left.\derf{}{\hat{\phi}^j(p)}
\derf{}{\hat{\phi}^k(p')}~\!\Gamma\matri{\phi,\ldots}\right|_{0}
=(2\pi)^4 \delta^{(4)}(p'+p)\,
\widetilde{\Gamma}^{\phantom{kj}}_{kj} (p',p)~\!,
\phantom{aaaaa}\label{Eq:Notacja1}
\end{eqnarray}
where  $ \cF(x)\equiv \int\,\volel{p} ~\! e^{-i p x} \hat{\cF}(p)$, 
and 0 denotes the stationary point.  
I have chosen the following parametrization of the 
complete 1PI 2-point functions in the  $A^\al_\mu$, $\phi^j$ and $h_\al$ 
sector; for vectors:     
\eqs{\label{Eq:Gamma2-vector} 
\widetilde{\Ga}^{\mu\nu}_{\al\be}(-q,q)
&\equiv&
-\eta^{\mu\nu}\Big[
q^2\id-M^2_V(q^2)
\Big]_{\al\be}
+q^\mu q^\nu\sL_{\al\be}(q^2)
\,,
}
for the scalar-scalar two point function I employ once again 
Eq. \refer{Eq:Gamma2-scalar}, and parametrize 
the mixed vector-scalar correlation functions  
 as  follows
\eq{\label{Eq:Gamma2-vec-sc}
\widetilde{\Ga}^{\mu}_{\al j}(-q,q)
\equiv
i\, q^\mu\, P_{\al j}(q^2)
=
-\widetilde{\Ga}^{\ \mu}_{j \al}(-q,q)
\,. 
}
Finally, the functions involving  Nakanishi-Lautrup multipliers have, 
to all orders of perturbation theory, the tree-level form 
\koment{str.Rxi2,por.MPW11}
\eqs{\nn
\widetilde{\Ga}^{\al\nu}_{\ \be}(-q,q)
&\equiv& i\,\de^{\al}_{\ \be}\, q^\nu
=-\widetilde{\Ga}^{\nu\al}_{\be}(-q,q)\,,
\nn\\[8pt]
\widetilde{\Ga}^{\al}_{\ j}(-q,q)
&=& -\,\U{\al}{j} = \widetilde{\Ga}^{\ \al}_{j}(-q,q)\,,
\nn\\[8pt]
\widetilde{\Ga}^{\al\be}(-q,q)
&\equiv&\xi^{\al\be}\,. 
}
For the reader's convenience, I list also the tree-level 
approximations to the full 1PI 2-point functions: \koment{str.Rxi45,por.MPW26}
\eqs{
\sL_{\al\be}(q^2) &=& \de_{\al\be}+\cO(\hb)\,,\qquad \qquad
\quad
M^2_V(q^2)_{\al\be} = (\cT_\al v)^\rmt (\cT_\be v) + \cO(\hb)\,,
\nn \\
P_{\al i}(q^2) &=& - \de_{ij} (\cT_\al v)^j + \cO(\hb)
\,,\qquad \,
M^2_S(q^2)_{ij} = \pa_{\cphi^i}\pa_{\cphi ^j} V^{GI}_{tree}(\cphi)|_{\cphi=v}+ \cO(\hb)\,,  \nn 
}
where $\cT_\al$ are real antisymmetric matrices forming the representation 
of the gauge Lie algebra on scalar fields $\phi^j$  
(note that, in this notation, the matrix elements of $\cT_\al$ 
contain the renormalized gauge coupling constants), 
$v$ is the all-order VEV of the scalar field $\cphi=\phi+v$ in the symmetric phase 
(of course, 
to the leading order, $v$ can be replaced with its $\hbar^0$ term), 
and $V^{GI}_{tree}(\cphi)$ is the gauge-invariant 
( $(\cT_\al\cphi)^j \pa_{\cphi ^j} V^{GI}_{tree}(\cphi) \equiv 0$ )
and independent of gauge-fixing parameters tree-level potential of 
Hermitian scalar fields. 

To find the connected propagators with (resummed) 
quantum corrections, one needs to solve the equation 
\eq{\label{Eq:Prop-Gen}
\widetilde{\Ga}_{IJ}(-p,p)\, \widetilde{G}^{JK}(p,-p)
=i\, \de_{I}^{\ K}\,,
} 
where $I$, $J$ and $K$ run over components of bosonic fields 
$\phi^n$, $A^\al_\mu$ and $h_\be$. 
I~have chosen the following parametrization of $\widetilde{G}^{JK}(p,-p)$; 
in the vector-vector block: \koment{str.Rxi45,MPW26}
\eq{\label{Eq:G2-vector} 
\widetilde{G}^{\be\de}_{\nu\rho}(q,-q)
=-i
\left[\eta_{\nu\rho}-\frac{q_\nu q_\rho}{q^2}\right]
\Big[ V(q^2)^{-1} \Big]^{\be\de}
+i\,\frac{q_\nu q_\rho}{q^2} \cA(q^2)^{\be\de}
\,,\
}
in scalar-scalar and vector-scalar blocks: \koment{str.MPW13} 
\eq{
\widetilde{G}^{\,jn}(q,-q) = i\, H^{jn}(q^2)\,,
\qquad 
\widetilde{G}^{\be n}_{\nu}(q,-q) = q_\nu E^{\be n}(q^2) = 
-\widetilde{G}^{n\be }_{\ \, \nu}(q,-q)\,,
}
the block that mixes the Nakanishi-Lautrup fields $h_\beta$ with vectors 
is written as \koment{MPW13}
\eqs{
\label{Eq:G2:hA}
\tG_{\!\be\, \rho}^{\ \,\de}(q,-q)
=-q_\rho J^{\de}_{\ \be}({q^2})
=-\tG_{\, \rho\be}^{\,\de }(q,-q)\,,
}
while the $h_\beta$-$\phi^n$ and $h_\be$-$h_\ga$ 
blocks can be parametrized as
\eq{\label{Eq:G2:h-h-phi}
\tG_{\be }^{\ \,n}(q,-q)
=i\,\,\cI(q^2)^{n}_{\ \be}=
\tG_{\, \ \be }^{\,n}(q,-q)
\,, 
\qquad 
\tG_{\be\ga }(q,-q)
= i\, K_{\be\ga}(q^2)\,.
}
Before discussing relations between 1PI two-functions originating 
from the Slavnov-Taylor identities of Becchi-Rouet-Stora symmetry (STids) \cite{BRS1,BRS2}, 
I~should stress that those relations are valid only at the stationary points 
of the 1PI generating functional $\Gamma[\ ]$. However, in certain 
applications, e.g. in order to calculate the 
effective potential of scalar fields, 
\footnote{See e.g. \cite{Martin2-new} for a recent 
determination of the 2-loop effective potential for an arbitrary renormalizable 
model in a subclass of gauges considered in the present paper.}
one needs the propagators in the presence of an arbitrary constant scalar 
background $\anti{\cphi}^j$ which differs from the VEV $v^j$. 
For this reason, I give also (in Appendix A)  
formulae for propagators which are completely generic 
solutions to Eqs. \refer{Eq:Prop-Gen}, and do not rely on the STids;  
those formulae are not needed in what follows -- it is in fact 
much simpler to solve \refer{Eq:Prop-Gen} once again in the presence of 
STids, instead of trying to simplify the results from the Appendix. 
Nevertheless, there are two relations, satisfied by a \emph{generic} solution 
to Eqs. \refer{Eq:Prop-Gen}, which are important in what follows 
\footnote{It is, perhaps, also worth saying that even though  
\refer{Eq:cAgeneric}-\refer{Eq:Egeneric} follow from \refer{Eq:Prop-Gen}, 
they can be also obtained independently of \refer{Eq:Prop-Gen}, as they express 
the non-renormalization 
theorem for gauge-fixing terms in the form appropriate for connected
(rather that 1PI) generating functional. \koment{por.str.Rxi37 i nastepne}
}
\koment{Rxi11}
\eq{\label{Eq:cAgeneric}
\cA(q^2)^{\al\de} = \U{\al}{j}E^{\de j}(q^2)
                    -\xi^{\al\be} J^{\de}_{\ \be}(q^2)\,,
}
\koment{Rxi56, cf.Rxi40}
\eq{\label{Eq:Egeneric}
q^2\, E^{\ga n}(q^2) =\U{\ga}{j} H^{jn}(q^2)-\xi^{\ga\al}\cI(q^2)^n_{\ \al}\,.  
}
Additionally, the transverse part of \refer{Eq:G2-vector} is also 
(\emph{independently} of STIds)
easy to express in terms of the 1PI two-point function \refer{Eq:Gamma2-vector}
\koment{str.Rxi45,MPW26: $V(q^2)\equiv-N(q^2)$}
\eq{\label{Eq:V}
V(q^2) = q^2\id-M^2_V(q^2)\,.
}

From now on, I assume that we are at the (physical -- see below) stationary 
point of the 1PI generating functional, so that STids \cite{BRS1,BRS2}
(
see also \cite{PiguetSorella} for an introduction, 
and \cite{ALLandau} for a derivation in the present notation
\footnote{Since I keep the $h_\al$ fields, the STids have exactly the 
same form as in the Landau gauge, thus the derivation 
from  \cite{ALLandau} applies in the present context.}
) 
yield relations between the 1PI 2-point functions; namely 
there exists matrices $B(q^2)^j_{\ \ga}$ and $\Om(q^2)^\al_{\ \ga}$
such that form-factors of 1PI 2-point functions satisfy the following constraints
\koment{por Rxi30, OLD:str.MPW12, por.MPW26,MPW71} 
\eq{
\label{Eq:STid1prime}
P_{\be j}(q^2)\,B(q^2)^j_{\ \ga}
=  
\left\{ q^2\, \sL_{\al\be}(q^2) + [M_V^2(q^2)-q^2\mathds{1}]_{\al\be} \right\}
\Om(q^2)^\al_{\ \ga}\,,
}
and
\eq{
\label{Eq:STid7prime}
q^2\,  P_{\al j}(q^2)\,\Om(q^2)^\al_{\ \ga} 
=  
S(q^2)_{ij}\, B(q^2)^i_{\ \ga}
\,.
}
In fact, there is a concrete prescription for calculating 
$\Om(q^2)$ and $B(q^2)$ in perturbation theory: let 
$\mathcal{L}^{s}$ be the BRS-exact part of the (tree-level) Lagrangian \cite{PiguetSorella}  \koment{(D:C.3)}
\begin{eqnarray}\nn 
\mathcal{L}^{s} &=&
s\!\left\{\overline{\omega}_\alpha [f^\al+\xi^{\al\be}h_\be/2]
\right\}
+ L_\alpha\, s({\omega}^\alpha)+K_i\, s(\phi^i)
+\bar{K}_a\, s(\psi^a)+K^\mu_\alpha\, s(A^\alpha_\mu)~\!,\phantom{aa}
\end{eqnarray}
where $s$ is the nilpotent BRS-differential \cite{PiguetSorella}, 
the first term on the right-hand-side contains (in addition to gauge-fixing 
terms 
from Eq.\refer{Eq:LagrGaugeFix}) the vertices involving 
ghost $\omega^\alpha$ and antighost $\overline{\omega}_\alpha$ fields, 
$\psi^a$ represents Majorana fields (i.e. natural counterparts of Hermitian 
bosonic fields),  while $K_i$, $\bar{K}_a$, $K^\mu_\al$ and $L_\al$ are 
external sources (antifields) 
controlling the  quantum corrections to BRS transformations \cite{BRS1,BRS2,PiguetSorella}. 
Then, the matrices $B(q^2)^j_{\ \ga}$ and $\Om(q^2)^\al_{\ \ga}$ 
\koment{str.MPW11} \kA
are given by the following derivatives at the stationary point: 
\eqs{\label{Eq:Antifields-Gh:Scalar}
\left.\derf{}{\hat{\om}^\ga(p)}
\derf{}{\hat{K}_i(q)}\Gamma\right|_{0}
&=&
(2\pi)^4 \delta^{(4)}(q+p)\, 
B(q^2)^i_{\ \ga} \,\,,
 \\[8pt]
\label{Eq:Antifields-Gh:Vector}
\left.\derf{}{\hat{\om}^\ga(p)}
\derf{}{\hat{K}^\mu_\al(q)}\Gamma\right|_{0}
&=&
(2\pi)^4 \delta^{(4)}(q+p)
\left\{i\,q_\mu\, {\Om}(q^2)^{\al}_{\ \ga}\right\}\,.
}
At the lowest order one finds \koment{str.Rxi45}
\eq{\label{Eq:B-Om-tree}
B(q^2)^i_{\ \ga}=(\TS_\ga v)^i+\cO(\hb)\,,
\qquad 
{\Om}(q^2)^{\al}_{\ \ga} = -\de^{\al}_{\ \ga} +\cO(\hb) \,,
}
in particular STids can be easily verified at the tree-level. 
Of course, at higher orders, the calculation of diagrams with external lines of ${K}^\mu_\al$, 
etc. is no different from calculations of
diagrams involving the propagating fields ${A}^\mu_\al$, etc.
(apart from simplifications due to vanishing 
propagators). 

Strictly speaking, STids \refer{Eq:STid1prime}-\refer{Eq:STid7prime} 
are satisfied only if \koment{Rxi29} $\langle h_\al\rangle=0$  
at the stationary point of $\Gamma[\ ]$; 
however in practice one can always find stationary points obeying this condition 
(see e.g. \cite{ALNielsen} and references therein), in particular 
equation $\langle h_\al\rangle=0$ can be interpreted as the Dashen's vacuum 
realignment condition (see e.g. \cite{WeinT2}) for the effective potential
$V_{eff}(\phi,h)$ that exhibits explicit breaking of the symmetry under 
global gauge transformations. In fact, 
the stationary points violating $\langle h_\al\rangle=0$ are inherently 
unphysical, as they lead to spontaneous breaking of the BRS symmetry 
and thus violate the quartet mechanism of Kugo and Ojima \cite{KugoOjima}, 
that is necessary for decoupling of unphysical modes from the physical $S$ 
matrix. 

I need just one additional identity to invert the matrix of 1PI 2-point functions; 
in linear gauges the ghost-antighost 1PI 2-point function is not independent 
of matrices $B(q^2)^i_{\ \ga}$  and ${\Om}(q^2)^{\al}_{\ \ga}$ appearing in the STids (see e.g. \cite{PiguetSorella}); in particular, the ghost-antighost propagator  $\widetilde{\cG}(q^2)^{\be}_{\ \al}$ satisfies (to all 
orders) \koment{Rxi31}
\eq{\label{Eq:G2:gh-antigh}
i[\widetilde{\cG}(q^2)^{-1}]^{\al}_{\ \be}
=-{q^2}\Om(q^2)^{\al}_{\ \be}+\U{\al}{j}B(q^2)^{j}_{\be}
\,.}
Clearly, $\widetilde{\cG}(q^2)$ has only unphysical 
poles, and therefore one expects that it can be useful for extraction of 
unphysical poles from  propagators of bosonic fields. 
In fact, as the first STids-triggered simplification, one finds that the 
mixed propagators involving the $h_\be$ fields are simply given by 
(cf. Eqs. \refer{Eq:G2:hA}-\refer{Eq:G2:h-h-phi})
\koment{str.Rxi47,Rxi54}
\eq{\label{Eq:h-prpp}
J^\be_{\ \de}(q^2) = i\, \Om(q^2)^\be_{\ \ga} \, 
   \widetilde{\cG}(q^2)^{\ga}_{\ \de}\,,\qquad\qquad  
\cI^j_{\ \de}(q^2) = i\, B(q^2)^j_{\ \ga} \, 
   \widetilde{\cG}(q^2)^{\ga}_{\ \de}\,,
}
\koment{str.Rxi55:}
and that $K_{\al\be}(q^2)=0$, i.e. the $h_\al$-$h_\be$ 
propagator vanishes. Thus, the asymptotic field associated with $h_\al$ 
describes zero-norm states, which have non-vanishing scalar products 
only with unphysical bosonic states \cite{KugoOjima}.

To express the scalar-scalar propagator $H^{jn}(q^2)$ in a simple form,
I need to define the following (non-symmetric!) matrix (cf. Eq. \refer{Eq:Gamma2-scalar}) \koment{str.Rxi57}
\eq{\label{Eq:T} 
T(q^2)_{ij} = S(q^2)_{ij}-P_{\al i}(q^2)\, \U{\al}{j}\,,
}
then (using the matrix-multiplication)  \koment{str.Rxi71}
\eq{\label{Eq:H-formT}
H(q^2) = T(q^2)^{-1}\,S(q^2)\, [T(q^2)^{-1}]^{\rmt} 
 	-\xi^{\al \de}\,\, \cI_{\al}(q^2)\, \cI_{\de}(q^2)^{\rmt}\,,
}
where $\cI_{\de}$ is a vector of  $\cI^j_{\ \de}$ propagators from 
Eq. \refer{Eq:h-prpp}. 
Note that $H(q^2)$ is explicitly symmetric; 
nonetheless the STids allows us to rewrite $H(q^2)$ in a form that is even 
more useful for extraction of physical poles. Using the relation 
\koment{str.Rxi72,Rxi75} 
\eq{\label{Eq:T-and-si}
T(q^2)  =  \frac{1}{q^2} S(q^2) \si(q^2)\,,
\quad {\rm where} \quad 
\si(q^2)^{i}_{\ j}  \equiv q^2\, \de^{\,i}_{ j} - C^{\,i}_{\ \al}(q^2)\, \U{\al}{j}   \,,
}
with
\eq{\label{Eq:Def-tC}
{C}^{j}_{\ \ga}(q^2)\equiv B(q^2)^{j}_{\ \be}\,[\Om(q^2)^{-1}]^{\be}_{\ \ga}
\,,
}
one gets \koment{strRxi72}
\eq{\label{Eq:H-formSigma}
H(q^2) = (q^2)^2\, \si(q^2)^{-1}\,S(q^2)^{-1}\, [\si(q^2)^{-1}]^{\rmt} 
 	-\xi^{\al \de}\,\, \cI_{\al}(q^2)\, \cI_{\de}(q^2)^{\rmt}\,.
}
Thus every massive root of $\det(S(q^2)) =0 $ yields a singularity of the 
scalar-scalar propagator, what justifies my earlier claim  that the 1PI correlation 
functions in the presence of $h_\al$ fields are more physical than their 
counterparts obtained by integrating $h_\al$'s out. I will have more to say 
about the structure  of poles later on, but first I will complete the list of 
propagators.  The mixed vector-scalar propagator has the form 
\koment{strRxi58,Rxi68,Rxi15}
\eq{\label{Eq:E}
E^{\de n}(q^2) = \cW(q^2)^\ga_{\ \, i}\, [T(q^2)^{-1}]^{ni}\, 
	J^{\de}_{\ \ga}(q^2)\,,
}
where 
\eq{\label{Eq:W}
\cW(q^2)^\ga_{\ \, j} \equiv \U{\ga}{j}-\xi^{\ga \al} P_{\al j}(q^2)\,,
}
while $J^{\de}_{\ \ga}$ is the propagator from Eq. \refer{Eq:h-prpp}. 
Note that $\cW(q^2)$ is momentum independent \emph{only} at the tree-level; 
the standard $R_\xi$ gauges (see e.g. \cite{WeinT2} ) are defined by the 
 `t Hooft's condition $\cW(q^2) = \cO(\hb)$, and thus the scalar-vector mixing 
reappears in the $R_\xi$ gauges at the quantum level. 

Last but not least, one gets the scalar part of the vector-vector propagator 
\refer{Eq:G2-vector} (recall that the transverse part is given by \refer{Eq:V} 
\emph{independently} of STids)
\koment{str.Rxi64}
\eq{\label{Eq:cA}
\cA(s)^{\al\de} = \frac{1}{s}\, \U{\al}{j}\, \U{\de}{n} \, H^{nj}(s)
+\frac{1}{s} \xi^{\al\de} 
-\xi^{\de\be} \Om(s)^{\al}_{\ \ep}\, i\, \widetilde{\cG}(s)^{\ep}_{\ \be}
-\xi^{\al\be} \Om(s)^{\de}_{\ \ep}\, i\, \widetilde{\cG}(s)^{\ep}_{\ \be}
\,.
} 
where $s\equiv q^2$ while $\widetilde{\cG}(s)$ is the ghost-antighost 
propagator. 

The above formulae are valid to all orders of perturbation theory; 
nonetheless they have an amusing consequence at the tree-level: 
all the {tree-level} propagators, 
in contrast to the $R_\xi$ gauges, 
are first order polynomials in $\xi$. 
Indeed, masses of unphysical modes are produced entirely by the $\U{\al}{i}$ 
parameter which (in the $R_\xi$ gauges)  becomes a function of $\xi$, once 
the `t Hooft's fine-tuning condition $\cW(q^2)=\cO(\hbar)$ is imposed. 
In particular, the $\U{\al}{i}$ parameter cures the $1/q^4$ IR-singularities 
which appear in the second term of the scalar-scalar propagator 
\refer{Eq:H-formSigma} for $\U{\al}{i}=0\neq \xi^{\al\be}$ (of course, since $\cI_\al = 0$ in the unbroken phase, these singularities exists only 
in spontaneously broken gauge theories).  

The behavior of the scalar-scalar propagator  $i\,H^{jn}(q^2)$ about the massive 
roots $q^2 = m^2_{S(\ell)} \neq 0$ of $\det(S(q^2))=0$ follows immediately 
from Eq. \refer{Eq:H-formSigma}. I use the parametrization of $S(q^2)$
given in 
\refer{Eq:Gamma2-scalar}. Suppose that we have the eigenvectors 
$\tilde{\ze}^k_{S[\ell_r]}$ obeying Eqs. 
\refer{Eq:xi-Eig-GENERAL-scalars}-\refer{Eq:norm-cond-GENERAL-scalars}; 
defining 
\eq{\label{Eq:ze_S}
{\ze}^k_{S[\ell_r]} = m^2_{S(\ell)}\, [\si(m^2_{S(\ell)})^{-1}]^{k}_{\ j}\, 
													\tilde{\ze}^j_{S[\ell_r]} \,,
}
one gets (cf. Eq. \refer{Eq:D-as-GENERAL-scalars-indi})
\eq{
\widetilde{G}^{kj}(q,-q)
\approx 
\sum_r
\, 
{\ze}^k_{S[\ell_r]} 
\,
\frac{i}{q^2-m^2_{S(\ell)}}
\,
{\ze}^j_{S[\ell_r]}
\,, \qquad {\rm for} \qquad 
q^2 \approx m^2_{S(\ell)} \neq 0\,. 
}
In theories without physical massless scalars, 
all the remaining poles of the scalar-scalar propagator are unphysical
(see the discussion in Appendix B). The only naturally massless scalars 
are Goldstone bosons of nonlinearly realized exact global symmetries; 
since these days such symmetries are quite commonly considered as inconsistent 
with quantum gravity, I will not give here the generic prescription 
for the ${\ze}^k_{S[\ell_r]}$-like vectors associated with them; nonetheless 
such a prescription can be obtained by studying the first term in Eq. 
\refer{Eq:H-formSigma}  
more carefully, once a concrete model is chosen. It is, however,  worth saying  
that a simple and \emph{generic} prescription for ${\ze}^k_{S[\ell_r]}$ vectors 
corresponding to the physical Goldstone bosons exists in the Landau gauge 
\cite{ALLandau}, and shows that their directions are fixed by the spontaneously 
broken gauge symmetry rather than the underlying global symmetry.

Thus, excluding theories with \emph{physical} massless scalars, 
the asymptotic field $\vphi^j$ corresponding to the renormalized scalar field $\phi^j$ has 
the form analogous to  \refer{Eq:AsymField}
\eq{\label{Eq:AsymField-scalar-GT}
\vphi^j= 
\sum_{  \ell } {}^\prime \sum_r
{{\zeta}}^j_{S[\ell_r]}  
\Phi^{\ell_r} 
+ \ldots
\,,
}
where $\Phi^{\ell_r}$ are canonically normalized free scalar fields 
corresponding to real pole masses (as indicated by the prime 
on the sum over $\ell$), while the ellipsis indicates the contributions 
of unphysical asymptotic states. (As is clear from the discussion above,  
even in the presence of physical massless scalars, Eqs. 
\refer{Eq:AsymField-scalar-GT} and \refer{Eq:ze_S} give the 
contributions of physical \emph{massive} scalars to the asymptotic scalar fields 
$\vphi^j$.)

The reader might be worried that the prescription \refer{Eq:ze_S} for the 
${{\zeta}}^j_{S[\ell_r]}$ coefficient depends, through the $\si(m^2_{S(\ell)})$ 
matrix, 
on the 1PI correlation functions of antifields 
\refer{Eq:Antifields-Gh:Scalar}-\refer{Eq:Antifields-Gh:Vector}. 
I could argue that, in principle, the $\sigma(q^2)$ function can be expressed  
in terms of $S(q^2)$ and the mixed scalar-vector 1PI two-point function 
$P_{\al i}(q^2)$, cf. Eqs. \refer{Eq:T-and-si} and \refer{Eq:T}, 
although obtaining $\si(m^2_{S(\ell)})$ in this way necessarily involves taking the limit  $q^2\to m^2_{S(\ell)}$.
In practice, however, one should notice that the number of diagrams 
contributing to the correlation functions of antifields is always significantly 
smaller than the number of diagrams contributing to $P_{\al i}(q^2)$. 
Therefore the form \refer{Eq:ze_S}  of the prescription for ${{\zeta}}^j_{S[\ell_r]}$ is actually quite convenient, 
as it is (owing to the STids!) completely independent of 
$P_{\al i}(q^2)$, as well as of the scalar part $\cA^{\al\de}(q^2)$ of the vector-vector two point functions. In fact, the same is true for 
the prescription that gives us the asymptotic vector fields (see below).  
Thus, the functions $P_{\al i}(q^2)$ 
and $\cA^{\al\de}(q^2)$ are not needed in practice. 

Using the simple form of the propagators listed above, it 
is also easy to find the asymptotic vector fields $\vA^\al_{\mu}$  associated with the renormalized vector fields $A^\al_{\mu}$ 
\koment{str.Rxi104}
\eq{\label{Eq:AsymField-vec}
\vA^\al_{\mu}
 = 
\sum_{  \la } {}^\prime  \sum_r
{\zeta}^\al_{V[\la_r]}  
\bA^{\la_r}_{\mu} 
+
\U{\al}{j}
\sum_{  \ell } {}^\prime \sum_r
\frac{1}{m^2_{S(\ell)}} {{\zeta}}^j_{S[\ell_r]}  
\pa_\mu\Phi^{\ell_r} 
+ \ldots\,,
}
as before the ellipsis represents the contributions 
of unphysical asymptotic states. The first term in 
\refer{Eq:AsymField-vec} represents the spin-1 operators 
inferred from the transverse part of the vector-vector propagator 
\refer{Eq:G2-vector}; in particular the ${\zeta}^\al_{V[\la_r]}$ coefficients 
and the pole masses $m_{V(\la)}^2$ are obtained by the following 
replacements in Eqs. \refer{Eq:GapEq-0-scalar}-\refer{Eq:norm-cond-GENERAL-scalars}
\eq{\nn 
M^2_S(q^2) \mapsto M^2_V(q^2)\,, \qquad 
m_{S(\ell)}^2  \mapsto m_{V(\la)}^2\,, \qquad 
\tilde{\ze}_{S[\ell_r]} \mapsto {\zeta}^\al_{V[\la_r]},
}
where $M^2_V(q^2)$  parametrizes the 1PI two-point function of vector fields 
\refer{Eq:Gamma2-vector}. 
The coefficient $\bA^{\la_r}_{\mu}$ in Eq. \refer{Eq:AsymField-vec}
is a free Hermitian 
vector field of mass $m_{V(\la)}$ (in the unitarity gauge for 
$m_{V(\la)}\neq0$, or the Coulomb gauge for $m_{V(\la)}=0$) with canonically 
normalized propagator; the states 
created/annihilated by $\bA^{\la_r}_{\mu}$ and 
$\bA^{\la'_{r'}}_{\mu}\neq \bA^{\la_r}_{\mu} $ are orthogonal 
to each other, and the prime on the sum over $\la$ 
has the same meaning as for scalars. Note that the form of the second term 
in \refer{Eq:AsymField-vec}, which involves only on the 
objects present already in Eq. \refer{Eq:AsymField-scalar-GT},  follows 
unambiguously from the representation \refer{Eq:Egeneric} of the vector-scalar 
propagator, as well as (up to a sign) from the scalar part \refer{Eq:cA}
of the vector-vector propagator. \footnote{To arrive at this conclusion it is 
enough  
to realize that all the terms 
in Eqs. \refer{Eq:Egeneric} and \refer{Eq:cA} except the first ones do not have 
physical poles (cf. Eqs. \refer{Eq:h-prpp}). } 
Using the generic form (given in \cite{ALLandau}) of (unphysical) 
quantum fields having propagators with higher-order poles, one can also verify 
that unphysical states in Eqs. 
\refer{Eq:AsymField-scalar-GT}-\refer{Eq:AsymField-vec} form 
(together with their Faddeev-Popov counterparts) the standard quartet 
representations of the asymptotic BRS operator \cite{KugoOjima}, 
\footnote{
This can be done in an analogous way to the Landau gauge case, 
for which a detailed and completely generic proof can be found 
in \cite{ALLandau}. 
} 
thereby ensuring unitarity of transition amplitudes between physical states. 

Eq. \refer{Eq:AsymField-vec} shows, in particular,  that the amputated 
correlation functions 
of renormalized vector fields $A^\al_{\mu}$ produce nontrivial 
contributions to amplitudes of scalar particles created by $\Phi^{\ell_r}$, 
just as the correlation functions of scalars $\phi^j$ do. 
It is also easy to check directly from Eq. \refer{Eq:ze_S} that, 
in the special case of $R_\xi$ gauges, the coefficient 
$\U{\al}{j} {{\zeta}}^j_{S[\ell_r]}$ 
vanishes at the tree level. 
\footnote{In the $R_\xi$ gauges, $\U{\al}{j}$ is a (transposition of) null 
eigenvector of the tree-level contribution to $M_S^2(0)$, as follows from the 
STids; thus $\U{\al}{j}$ is orthogonal to the eigenvector 
$\tilde{{\zeta}}^j_{S[\ell_r]}$ corresponding to a nonzero mass, 
and then Eq. \refer{Eq:ze_S} shows that ${{\zeta}}^j_{S[\ell_r]}=\tilde{{\zeta}}^j_{S[\ell_r]}+\cO(\hb)$. 
\koment{Rxi97}.
}
Nonetheless, even in the $R_\xi$ gauges, 
non-vanishing contributions to $\U{\al}{j} {{\zeta}}^j_{S[\ell_r]}$  
are generated by the quantum corrections. Therefore   
Eqs. \refer{Eq:AsymField-vec} and  \refer{Eq:ze_S}, 
may come in handy even in this special case. 

\section{Conclusions}\label{Sec:Concl}

I gave explicit formulae for 
the propagators of a generic spin-1 gauge field theory model, 
valid to all orders of perturbation theory in the presence of 
mixing between vectors and scalars. 
Due to relations originating from the BRS symmetry, 
the propagators take a  marvelously simple form, 
which is a convenient starting point for the study 
of their behavior about the physical (as well as unphysical) poles. 

This way, I have arrived at Eqs. 
\refer{Eq:AsymField-scalar-GT} and 
\refer{Eq:AsymField-vec} which, accompanied by the explicit 
prescription for the relevant coefficients, in particular Eq. \refer{Eq:ze_S}, 
generalize the Lehmann-Symanzik-Zimmermann algorithm 
to the case of an arbitrary mixing between bosonic fields in gauge theories. 
The treatment of spin-1/2 fermions in this framework can be found in 
\cite{ALFermions}

\appendix 

\section{ Generic solution to Eq. \refer{Eq:Prop-Gen} }
In this section I give, for completeness, the generic solution to 
Eqs. \refer{Eq:Prop-Gen}, which does not rely on the STids and therefore is 
valid not only at the stationary points of the 1PI effective action $\Ga[\ ]$, 
but also,  for instance, in the presence of an arbitrary constant scalar 
background. To this end, I need (in addition to \refer{Eq:T} and \refer{Eq:W}) the following
combinations of form-factors of 1PI 2-point functions 
\koment{str.Rxi15,Rxi10,MPW11,MPW26}, 
\eq{
\sX_{\al\be}(q^2) =
q^2\sL_{\al\be}(q^2)
+M^2_V(q^2)_{\al\be}
-q^2 \de_{\al\be}\,,
}
\eq{
Q_{\ga i}(q^2) = q^2 P_{\ga i}(q^2) -\,\U{\be}{i}\,\sX_{\be\ga}(q^2)\,,
}
and 
\eq{
\De^{\al}_{\ \be}(q^2) = 
q^2\, \de^{\al}_{\ \be} -
\xi^{\al\ga}\sX_{\ga\be}(q^2)\,. 
}
The scalar-scalar propagator $H(q^2)^{ij}$ is given by 
the inverse of the following matrix (the order of indices matters!) 
\koment{str.Rxi20,Rxi57}
\eq{
[H(q^2)^{-1}]_{ij}
= T(q^2)_{ji} - 
Q_{\ga i}(q^2) [\De(q^2)^{-1}]^{\ga}_{\ \al}\,\cW^{\al}_{\ j}(q^2) \,,
}
the vector-scalar propagator reads 
\koment{str.Rxi20}
\eq{
E^{\be n}(q^2) =[\De(q^2)^{-1}]^{\be}_{\ \al}\,\cW^{\al}_{\ j}(q^2) H(q^2)^{jn}
\,,
}
the form-factor of Eq. \refer{Eq:G2:hA} 
reads \koment{str.Rxi20}
\eq{
J^{\be}_{\ \de}({q^2})
= 
[\De(q^2)^{-1}]^{\be}_{\ \de}
+Q_{\ka j}(q^2)E^{\be j}(q^2)
[\De(q^2)^{-1}]^{\ka}_{\ \de}
\,,
}
the scalar part \koment{Rxi11}
$\cA(q^2)^{\al\de}$ 
of the vector-vector propagator is now a linear combination 
of $J^{\de}_{\ \be}({q^2})$ and $E^{\be n}(q^2)$ as given by  
Eq. \refer{Eq:cAgeneric}, while the transverse part is given by \refer{Eq:V}. 
For completeness, I give here also the remaining two propagators 
that involve Nakanishi-Lautrup fields 
\koment{str.Rxi11}
\eq{
\cI(q^2)^{j}_{\ \ep} = P_{\ep i}(q^2) H^{ij}(q^2)
							 -\sX_{\ep\ga}(q^2)E^{\ga j}(q^2)\,,
}
and \koment{str.Rxi12}
\eq{
K_{\be\al}(q^2) = P_{\be j}(q^2) \cI(q^2)^{j}_{\ \al }
					   -\sX_{\be\ga}(q^2)J^{\ga}_{\ \al}(q^2)\,.
}
Two comments are in order. Firstly, matrices $K_{\be\ga}(q^2)$, 
$\cA^{\be\ga}(q^2)$ and  $H^{ij}(q^2)$ are indeed symmetric, even 
though it requires some patience to verify this. Secondly, 
as is clear from the formulae given in the main part of the paper, 
\emph{none} of the propagators corresponding to the actual stationary point 
of $\Gamma[\ ]$ 
has a pole at $q^2$ such that $\det(\De(q^2))=0$.


\section{Unphysical poles}

It is worth to take a closer look at the structure of the unphysical 
poles in the propagators of vector and scalar fields (see also \cite{BBBC}). 
To this end, in addition to STids 
\refer{Eq:STid1prime}-\refer{Eq:STid7prime}, it is important to keep  
in mind that 1PI correlation functions in 4 dimensions do not have poles  at 
finite orders of perturbation theory. 
Therefore, Eqs. \refer{Eq:h-prpp} show that the form-factors  $J(q^2)$ and $\cI(q^2)$ 
have only poles for $q^2 = m_{gh(\la)}^2$ where $m_{gh(\la)}^2$ is 
one of the poles of the ghost-antighost propagator. Then Eq. \refer{Eq:H-formT} 
shows that, apart from poles at $q^2 = m_{gh(\la)}^2$, the scalar-scalar 
propagator has only poles for $q^2$ such that $\det(T(q^2))=0$. 
Finally, Eqs. \refer{Eq:E} and \refer{Eq:cAgeneric} show that the same 
it true for the scalar-vector propagator and the scalar part of the 
vector-vector propagator. In other words, it is enough to study $T(q^2)$.

At the tree-level 
\eq{\label{Eq:T-exp}
T(q^2) = q^2\mathds{1} - \tau +\cO(\hb)\,,
}
with a momentum independent (but non-symmetric) matrix $\tau$;  
$\tau$ has at most $N_S$ eigenvectors, where $N_S$ is the number of scalar 
fields $\phi^j$ (real non-symmetric matrices are not diagonalizable in general). 
Thus, in perturbation theory, $T(q^2)$ has at most $N_S$ poles (where a pole 
with degeneracy is counted as multiple poles, as usual). 
Let us try to find as many eigenvectors of $\tau$ as possible. Eq. 
\refer{Eq:T-and-si} implies that every vector \refer{Eq:ze_S} 
obeys 
\koment{str.Rxi90} 
$T(m^2_{S(\ell)})  {\ze}_{S[\ell_r]}  = 0$, as long as $m^2_{S(\ell)}\neq0$, 
to all orders of perturbation theory.
This gives us  
$N_S-N_{\ker}$ eigenvectors of $\tau$ at the tree-level, where $N_{\ker}$ 
is the number of null eigenvectors of $M_S^2(0)$ or (equivalently) $S(0)$.  
The STid \refer{Eq:STid7prime} shows that, to all orders of perturbation theory, 
every vector $C^{j}_{\ \ga}(0)$, cf. Eq \refer{Eq:Def-tC}, 
belongs to the kernel of $S(0)$. Of course, not all of $C^{j}_{\ \ga}(0)$ are 
linearly independent, but STids yield also the following equivalence 
(at every finite order of perturbation theory) 
\eq{\label{Eq:M_V-equiv}
M_V^2(0)_{\al\be}\La^\be=0
\qquad\quad 
\Leftrightarrow
\qquad\quad 
C^i_{\ \be}(0)\,\La^\be(0)=0\,, 
} 
(see \cite{ALLandau}, Eq. (150)), 
i.e. the number of linearly independent vectors $C^i_{\ \be}(0)$ is 
the same as the number of non-massless gauge bosons (what is obvious at the 
tree-level). Therefore, in models without physical massless scalars, 
$N_{\ker}$ is imply the number of massive gauge bosons. 
\koment{Rxi83} 
It is interesting to note that Eq. \refer{Eq:G2:gh-antigh} 
now implies that \footnote{In what follows, with a little abuse of notation, $[\widetilde{\cG}(q^2)^{-1}]$ represents  
the 1PI 2-point functions of ghosts, regardless of whether it is an invertible matrix or not.} 
\eq{
[\widetilde{\cG}(0)^{-1}]^{\al}_{\ \be}\,[\Om(0)^{-1}]^{\be }_{\ \ga}\La^\ga = 0\,,
}
where $\La^\ga$ obeys one of the conditions in \refer{Eq:M_V-equiv}, 
and thus for every massless gauge boson there is a massless ghost field 
(recall that $\Om(q^2)$ is invertible at the tree-level and therefore 
it is also invertible at every finite order). 
Of course, the inverse theorem is, in general, not true (all ghosts are 
massless if $\,\U{\al}{i}=0$). Nonetheless, since the Landau-gauge case was 
studied in details in \cite{ALLandau} and since non-Landau gauges 
with massless ghosts corresponding to broken gauge symmetries have 
severe IR divergences (cf. the second term in \refer{Eq:H-formT}), 
it is reasonable to assume that $\U{\al}{i}$ has been chosen in such a way 
that all massless ghosts correspond to massless gauge bosons.  
Additionally, I will assume the total number of independent vectors $\Th_{gh(\la_r)}$ obeying 
\eq{\label{Eq:cond}
[\widetilde{\cG}(m_{gh(\la)}^2)^{-1}]\, 
[\Om(m_{gh(\la)}^2)^{-1}]\, 
 \Th_{gh(\la_r)} = 0\,, 
\quad{\rm with} \quad m_{gh(\la)}^2 \neq 0 \,,
}
is equal to the number of massive ghosts (and thus, to the number of massive gauge bosons). 
\footnote{This is  
merely a technical assumptions: if it is not satisfied, then  one has  to 
study the relations between the generalized (rather than plain) eigenvectors of $[\widetilde{\cG}(m_{gh(\la)}^2)^{-1}]$ and $T(m_{gh(\la)}^2)$, see e.g. 
\cite{ALFermions} and references therein. 
}
This ensures that the propagator $\widetilde{\cG}(q^2)$ has only simple poles 
in perturbation theory  
(note that the tree-level mass matrix of ghosts is not symmetric, 
and therefore this is not guaranteed even at the lowest order). 

Under these assumptions it is easy to find the 
missing $N_{\ker}$ eigenvectors of $\tau$ from \refer{Eq:T-exp}. 
Comparing \refer{Eq:G2:gh-antigh} 
with \refer{Eq:T-and-si} one gets (to all orders)
\koment{Rxi76}
\eq{
\si(m_{gh(\la)}^2)^i_{\ j} C^{j}_{\ \ga}(m_{gh(\la)}^2) \Th_{gh(\la_r)}^\ga = 0,  
} 
and thus $T(m_{gh(\la)}^2)_{ij} C^{j}_{\ \ga}(m_{gh(\la)}^2) \Th^\ga_{gh(\la_r)} = 0$ 
if  $m_{gh(\la)}^2 \neq 0$. 
Let's suppose that a linear combination of 
$C^{j}_{\ \ga}(m_{gh(\la)}^2) \Th_{gh(\la_r)}^\ga$ vanishes: 
\koment{str.Rxi87}
\eq{
\sum_{\la}\sum_r X^{(\la_r)} C^{j}_{\ \ga}(m_{gh(\la)}^2) \Th^\ga_{gh(\la_r)} = 0.
}
Contracting the above equation with $\U{\al}{j}$, 
using \refer{Eq:cond} and the explicit form \refer{Eq:G2:gh-antigh} 
of the ghost-antighost 2-point function we now get 
\eq{\label{Eq:eeee}
\sum_{\la}\sum_r  X^{(\la_r)} m_{gh(\la)}^2
 \Th^\ga_{gh(\la_r)} = 0.
}
Under the assumptions listed above, $\Th^\ga_{gh(\la_r)}$ are linearly 
independent and correspond to $m_{gh(\la)}^2\neq0$,
thus \refer{Eq:eeee} implies $X^{(\la_r)} =0 $, 
what proves that $C^{j}_{\ \ga}(m_{gh(\la)}^2) \Th^\ga_{gh(\la_r)}$ 
form the system of $N_{\ker}$ linearly independent vectors obeying  
$T(m_{gh(\la)}^2)_{ij} C^{j}_{\ \ga}(m_{gh(\la)}^2) \Th^\ga_{gh(\la_r)} = 0$.

To summarize, in theories without physical massless scalars, 
under a natural assumption that $\U{\al}{i}$ produces masses 
for all ghost that do not correspond to massless gauge bosons, 
the     
$T(q^2)^{-1}$ matrix has two kinds of poles: (1) physical poles at $q^2 = m^2_{S(\ell)} \neq 0$ 
where $m^2_{S(\ell)}$ is a massive pole of $S(q^2)^{-1}$, 
and (2) unphysical poles at $q^2 = m_{gh(\la)}^2 \neq 0$, 
where $m_{gh(\la)}$ is a pole mass of a \emph{massive} ghost.  
This statement is valid at every finite order of perturbation theory,
not just in the tree-level approximation.

\end{document}